\documentclass[english,gc, manuscript]{copernicus}
\usepackage[T1]{fontenc}
\usepackage{array}
\usepackage{multirow}
\usepackage{graphicx}
\usepackage{rotating}
\usepackage[authoryear]{natbib}

\makeatletter

\providecommand{\tabularnewline}{\\}

\DeclareRobustCommand{\disambiguate}[3]{#3}

\makeatother

\usepackage{babel}
\begin{document}
\title{Schools of all backgrounds can do physics research: On the accessibility
and equity of the PRiSE approach to independent research projects}
\author[1,{*}]{Martin O. Archer}
\affil[1]{School of Physics and Astronomy, Queen Mary University of London,
London, UK}
\affil[{*}]{now at: Space and Atmospheric Physics, Department of Physics, Imperial
College London, London, UK}
\correspondence{Martin O. Archer\\
(martin@martinarcher.co.uk)}
\runningtitle{Accessibility and equity of PRiSE}
\runningauthor{M.O. Archer}
\maketitle
\nolinenumbers
\begin{abstract}
Societal biases are a major issue in school students' access to and
interaction with science. Schools engagement programmes in science
from universities, like independent research projects, which could
try and tackle these problems are, however, often inequitable. We
evaluate these concerns applied to one such programme, `Physics Research
in School Environments' (PRiSE), which features projects in space
science, astronomy, and particle physics. Comparing the schools involved
with PRiSE to those of other similar schemes and UK national statistics,
we find that PRiSE has engaged a much more diverse set of schools
 with significantly more disadvantaged groups than is typical. While
drop-off occurs within the protracted programme, we find no evidence
of systematic biases present. The majority of schools that complete
projects return for multiple years of the programme, with this repeated
buy-in from schools again being unpatterned by typical societal inequalities.
Therefore, schools'  ability to succeed at independent research projects
appears independent of background within the PRiSE framework. Qualitative
feedback from teachers show that the diversity and equity of the programme,
which they attribute to the level of support offered through PRiSE's
framework, is valued and they have highlighted further ways of making
the projects potentially even more accessible. Researcher-involvement,
uncommon in many other programmes, along with teacher engagement and
communication are found to be key elements to success in independent
research projects overall.
\end{abstract}

\introduction{}

It has long been the case that the Science, Technology, Engineering,
and Mathematics (STEM) sectors have shown, both within higher education
and in the workforce, systemic biases against women, other non-male
genders, those from ethnic-minorities, and socially-disadvantaged
groups \citep[e.g.][]{case14}. Inequalities are present even at the
secondary/high school level, where students from under-represented
or disadvantaged backgrounds, despite being interested in science,
have fewer opportunities to engage with science both inside and outside
of school \citep{wellcome19}. These societal issues constitute major
inequities on young people that influence their opportunities, self-perception,
and ultimately subject/career choices. It is therefore important that
STEM engagement programmes aim to equitably include students from
these groups, taking into account all of the factors that might support
or prevent their engagement.

Independent research projects are opportunities enabling school students
to conduct open-ended investigations in science. While some independent
research project initiatives specifically target traditionally under-represented
groups (with many of these being in the USA), a recent global review
\citep{bennett16,bennett18} found that in general there are equity
issues relating to participation in such projects --- they are relevant
to all young people, but only a small minority are able to access
them. The authors note that despite emerging evidence that independent
research projects can result in improved engagement and attitudes
towards science amongst students from under-represented groups, further
work is required to more fully explore the potential benefits of independent
research projects on them.

`Physics Research in School Environments' (PRiSE) is a framework for
independent research projects for 14--18 year-old school students
that are based around cutting-edge physics research and mentored by
active researchers \citep{archer_report17,archer_prise_framework20}.
Unlike some citizen science initiatives with schools, which due to
their focus on answering specific science questions can sometimes
result in an inauthentic research experience focused around crowdsourcing
\citep{bonney09,bonney16,shah16}, PRiSE as a `research in schools'
programme was devised in an audience-focused way, with the benefits
to participants being of primary importance (discussed in more detail
in \citealp{archer_prise_framework20}).  Thus far the four PRiSE
projects summarised in Table~\ref{tab:projects} have been developed
at Queen Mary University of London (QMUL) since 2014 and the framework
is now being adopted by other institutions who are developing their
own projects applied to their specific areas of physics research.
The programme aims to equitably include significant numbers of students
from demographic groups which are under-represented in higher education
and STEM. Projects run from the start of the UK academic year in September
to just before the spring/Easter break in March, a duration of approximately
six months. The role of the teacher in these projects is chiefly one
of encouraging their students to persist, providing what advice they
can, and then communicating with the university. Teachers are not
expected to fully manage the projects, which is why numerous modes
of support are provided from active researchers who have the expertise
and skills in the areas of each project. This support offered to students
and teachers comes in the form of a suite of bespoke resources along
with the following intervention stages each year:
\begin{itemize}
\item \textbf{Assignment (Jun--Jul)} Teachers sign their school up for
a PRiSE project and are informed of the outcome before the summer
break.
\item \textbf{Kick-off (Sep--Oct) }An introductory talk and hands-on workshop,
either in-school or as an evening event on university campus.
\item \textbf{Visit (Dec--Feb) }Researchers visit the schools to mentor
students on their project work.
\item \textbf{Comments (Mar)} Researchers provide comments on students'
draft presentations near the end of the project.
\item \textbf{Conference (Mar)} Students present their project work as either
posters or talks at a student conference held on campus and attended
by teachers, family, and researchers.
\end{itemize}
and any further ad hoc communications as needed on an individual school
basis. Evaluation has shown that all of these elements of support
are almost equally important and necessary in the eyes of students
and teachers \citep{archer_prise_framework20}. This paper assesses
whether the approach taken and level of support provided by PRiSE
enables schools from all backgrounds to participate and succeed in
independent research projects. Section~\ref{sec:Participation} investigates
the diversity of schools that have participated in PRiSE, benchmarking
them against UK national statistics as well as schools involved in
other similar programmes. We then investigate retention of the schools
in PRiSE, both within each academic year and across multiple years,
in section~\ref{sec:Retention}. Finally, feedback from teachers
relating to diversity, accessibility and equity are presented in section~\ref{sec:Feedback}.
Impacts of the programme upon students, teachers and schools and whether
these are potentially affected by background are discussed in a companion
paper \citep{archer_prise_impact20}.

\begin{table*}
\begin{footnotesize}%
\begin{tabular}{>{\raggedright}p{0.25\columnwidth}lll>{\raggedright}p{0.33\columnwidth}}
\textbf{Project} & \textbf{Abbreviation} & \textbf{Years} & \textbf{Field} & \textbf{Description}\tabularnewline
\hline 
\noalign{\vskip6pt}
Scintillator Cosmic Ray Experiments into Atmospheric Muons & SCREAM & 2014--2020 & Cosmic Rays & Scintillator -- Photomuliplier Tube detector usage\tabularnewline
\noalign{\vskip6pt}
Magnetospheric Undulations Sonified Incorporating Citizen Scientists & MUSICS & 2015--2020 & Magnetospheric Physics & Listening to ultra-low frequency waves and analysing in audio software\tabularnewline
\noalign{\vskip6pt}
Planet Hunting with Python & PHwP & 2016--2020 & Exoplanetary Transits & Learning computer programming, applying this to NASA Kepler and TESS
data\tabularnewline
\noalign{\vskip6pt}
ATLAS Open Data & ATLAS & 2017--2020 & Particle Physics & Interacting through online tool with LHC statistical data on particle
collisions\tabularnewline
\end{tabular}\end{footnotesize}

\caption{A summary of the existing PRiSE projects at QMUL.\label{tab:projects}}
\end{table*}

\section{Participation\label{sec:Participation}}

As of March 2020, 67~schools have been involved in PRiSE. A full
list of (anonymised) schools is given in Appendix~\ref{sec:PRiSE-schools-table}.
Figure~\ref{fig:map} demonstrates that these schools (blue) have
been fairly broadly spread across Greater London rather than being
focused solely around Queen Mary (red). Schools targeting has been
limited to London to enable researchers to build relationships with
the schools via their in-person interactions throughout the 6~month
projects. Most schools have participated directly with Queen Mary,
though we note that some have been involved as a partnership of local
schools (there have been 8~partnerships across 22~schools, listed
in Appendix~\ref{sec:PRiSE-schools-table}). Such partnerships could
provide an additional support network to students and teachers as
well as making interventions more efficient for researchers. However,
we have found these partnerships to have been somewhat hit-or-miss
so far within PRiSE --- while kick-off events with all partner schools
present have typically worked, following this the schools have not
always worked with their partners on the projects. Further investigation
is required to understand what makes these partnerships work.

Here we evaluate the diversity of schools engaged in PRiSE. We limit
this analysis to publicly available data concerning the schools and
their local areas, and we did not collect any protected characteristics
(such as gender or race) or sensitive information (such as socio-economic
background) from the students involved. This was done for both ethical
and practical reasons, bearing in mind that this is a schools engagement
programme delivered and evaluated by physics researchers and not an
educational research project in and of itself. For example, it was
deemed that requiring students or their teachers to provide protected
or sensitive information upfront would have risked some students,
or indeed entire schools, declining to participate. This limits the
conclusions that can be made to only the school-level. However, it
has been recognised that the clustering of students within schools
results in students within the same school having more in common with
each other than with students in different schools, an important consideration
in the uptake of post-compulsory physics education for example \citep{gill11}.
While multilevel models could account for this hierarchy, this is
beyond the scope of what is practical for PRiSE. We note that schools
typically involve entire (or significant fractions of) cohorts of
A-level physics students in PRiSE and so while we have no indication
that PRiSE students differ in any substantive way from their schools'
wider student-base, we cannot rule out that they may not necessarily
be representative. Finally, since one of the aims of PRiSE is to impact
on teachers' practice and schools' STEM environments, school-level
considerations are valuable in this context regardless of the specific
characteristics of the students engaged in PRiSE.

We benchmark  school-level data against UK national statistics as
well as schools listed on the websites of two other similar UK-based
programmes of research-based physics projects for schools, \citet[$n=178$]{iris}
and \citet[$n=22$]{hisparc}. While we also looked at schools involved
with \citet{orbyts} who specifically mention targeting disadvantaged
groups, finding very similar results to PRiSE, with only 17~schools
listed there is limited scope for detailed statistical comparison
and so we have omitted this programme here. We make no comment on
the reasons behind the makeup of schools involved with different programmes,
since this would require specific qualitative research into how each
programme's provision model and targeting affects participation. Information
about schools was first obtained from the `Get information about schools'
database, formerly known (and henceforth referred to in this paper)
as Edubase \citep{edubase}. For more information on the UK schools
system, please see Appendix~\ref{sec:school-types}. While all PRiSE
and HiSPARC schools could be found in Edubase, only 154 of the listed
IRIS schools could be identified (based on UK postcodes).

\begin{figure*}
\begin{centering}
\includegraphics[width=0.8\columnwidth]{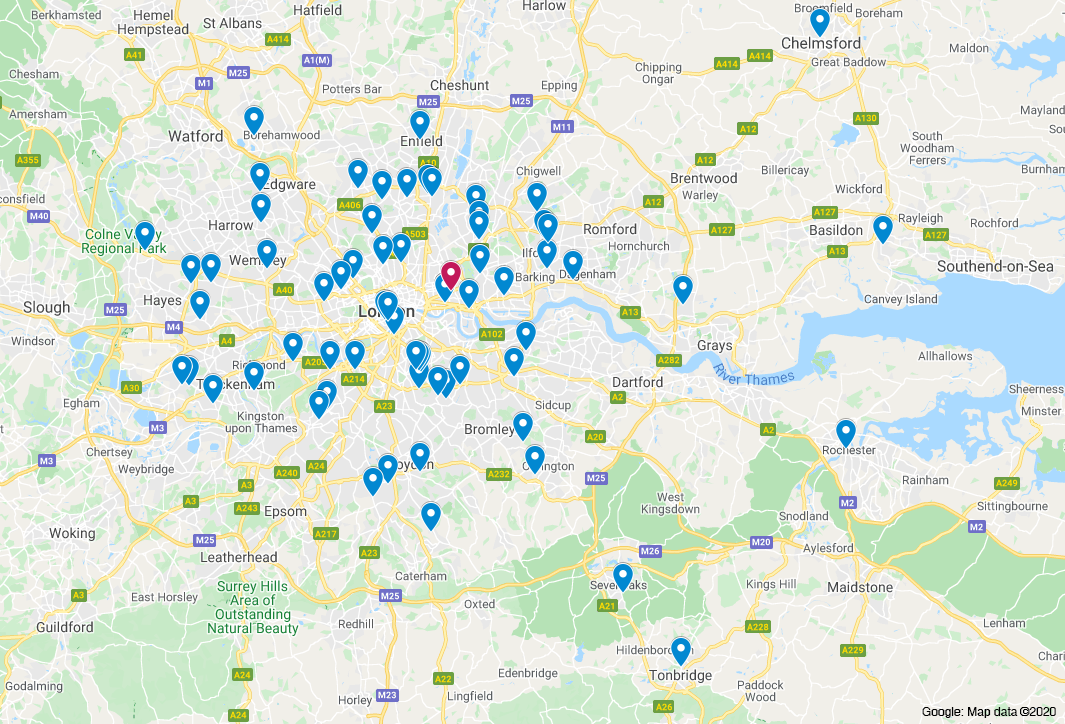}
\par\end{centering}
\caption{Map of all schools involved with PRiSE (blue) with Queen Mary University
of London's location shown in red.\label{fig:map}}
\end{figure*}

Figure~\ref{fig:school-types} shows the makeup of school categories
(explained in Appendix~\ref{sec:school-types}) from this database
across the three programmes, showing little overall difference between
them --- in a chi-squared test of independence $\chi^{2}\left(8\right)=8.45$
corresponding to $p=0.391$. While none of the differences between
the programmes are strictly statistically significant (using a difference
in binomial proportions test) due to the relatively small numbers
of schools compared to the population, it should be noted that IRIS
features proportionally more independent schools than PRiSE ($+0.10$
absolute and $1.46\times$ relative, $p=0.107$), HiSPARC involves
more academies than PRiSE ($+0.16$ and $1.40\times$, $p=0.260$),
and PRiSE works with more local authority maintained schools than
both HiSPARC ($+0.08$ and $1.52\times$, $p=0.418$) and IRIS ($+0.09$
and $1.60\times$, $p=0.147$) . However, it is clear that none of
the programmes are truly representative of all schools nationally
by category. This is also the case when looking at schools' admissions
policies (again see Appendix~\ref{sec:school-types} for further
background). The $10\pm4\%$ of selective schools in PRiSE is  more
than the $1\%$ nationally listed in Edubase, though we note that
both HiSPARC and IRIS feature even higher proportions of selective
schools than PRiSE at $14\pm9\%$ ($p=0.342)$ and $24\pm4\%$ ($p=0.007$)
respectively, where the uncertainties refer to the standard (i.e.
68\%) \citet{clopper34} confidence interval in a binomial proportion.
To address the imbalances in school categories and admissions policies
of PRiSE schools, as of 2019 we have implemented a policy that all
independent and selective schools must partner with local state or
girls' schools, including them in their project work. While this was
something which some independent schools had voluntarily been doing
previously, we have encountered some resistance to this policy by
a number of independent schools. Schools which refused the policy
were not allowed to participate, even if they had worked with us previously.
While several other schools agreed to the principle and tried to implement
partnerships, some failed to do so due to them not being able to draw
from existing local school partnerships, limited time from the application
to the summer holidays, and poor communication between teachers at
different schools. These schools were allowed to participate, with
the expectation that they put further effort in to establish these
partnerships ready for the next academic year, which they seemed willing
to do.

\begin{figure*}
\begin{centering}
\includegraphics[width=1\columnwidth]{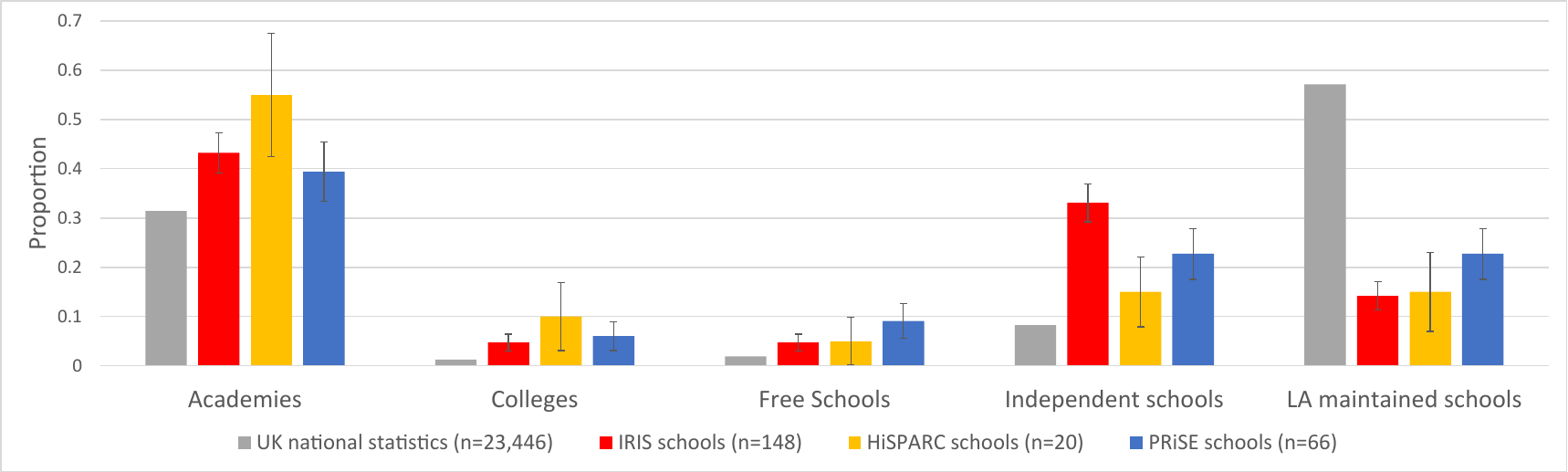}
\par\end{centering}
\caption{Distribution of school categories involved in IRIS (red), HiSPARC
(yellow) and PRiSE (blue) projects compared to UK national statistics
(grey). Error bars denote standard \citet{clopper34} intervals.\label{fig:school-types}}
\end{figure*}

\begin{figure*}
\begin{centering}
\includegraphics[width=1\columnwidth]{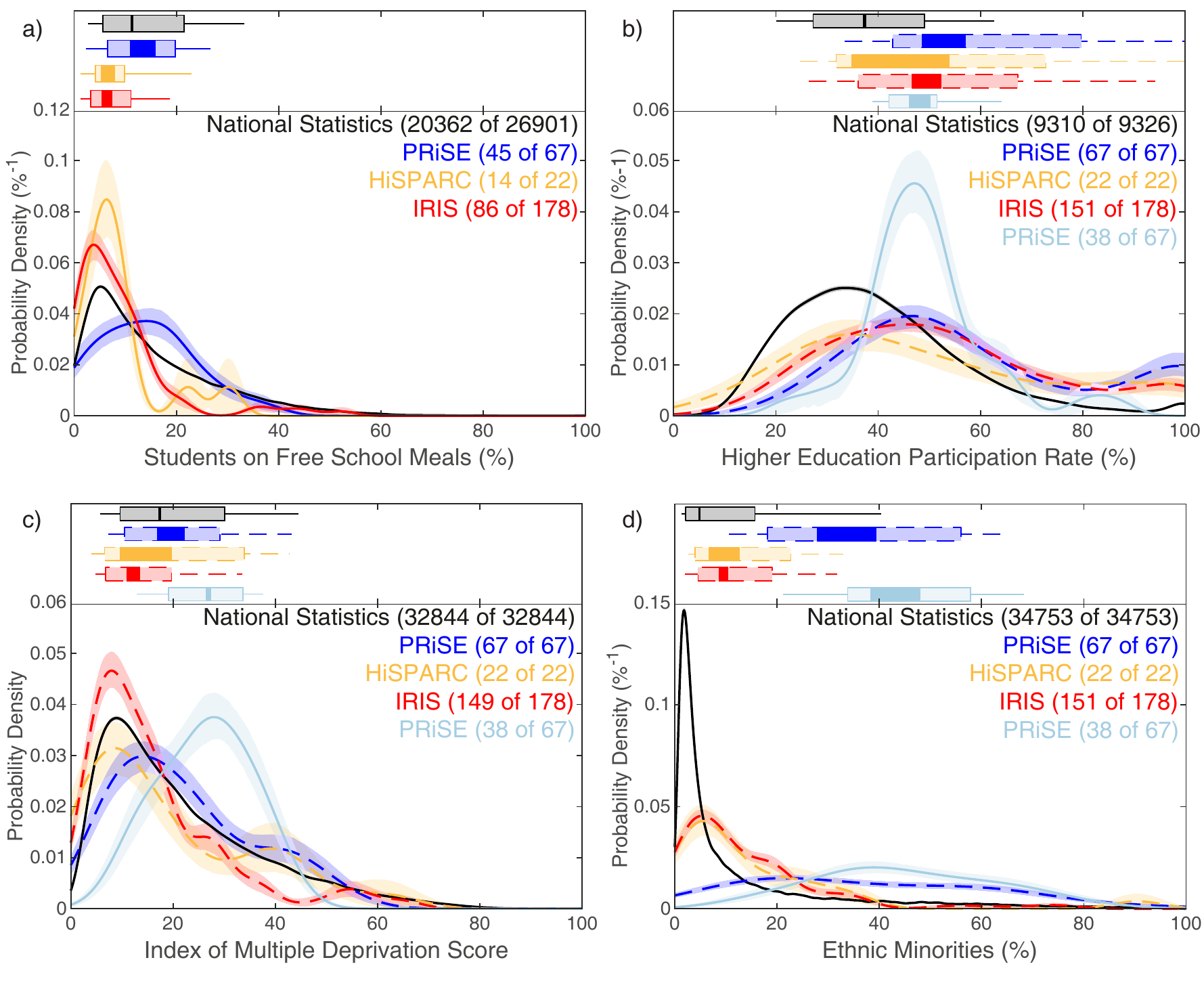}
\par\end{centering}
\caption{Distributions of a) students on free school meals, b) higher education
participation rate, c) index of multiple deprivation score, and d)
ethnic minorities. Boxplots in the top panels have whiskers covering
10--90\%, boxes spanning 25--75\%, and bands depicting the standard
confidence interval in the median. Lower panels show kernel density
estimates along with bootstrapped standard confidence intervals. Dashed
lines indicate only local data is used, which does not cover the schools'
full catchment areas.\label{fig:metrics}}

\end{figure*}

Beyond school type and admissions policy we look at several other
metrics for the backgrounds of the schools' students. We detail in
Appendix~\ref{sec:metrics-methods} how we have combined various
datasets in order to assess these. For PRiSE schools these methods
results in two different metric values for each school --- one covering
the school's full catchment area (the entire area from which they
draw students) and another purely pertaining to the school's local
census area (the immediate area surrounding the school's location).
We consider the full catchment area data to be more reflective of
a school's student base compared to the local data. For schools outside
of London, however, we only have access to the local data. Note that
the two types of data can result in rather different values for a
school despite the underlying distributions across all London schools
being similar (see Appendix~\ref{sec:metrics-methods} for further
discussion). In Figure~\ref{fig:metrics} distributions of the gathered
metrics are shown in two formats. Top panels display boxplots depicting
quantiles of the metric. Bottom panels depict kernel density estimates
of the continuous probability distributions, where Gaussian kernels
of optimal bandwidth from the \citet{silverman86} rule have been
applied to each dataset. Standard confidence intervals are estimated
by bootstrapping 1,000 random realisations of the data \citep{efron93},
taking quantiles of their resulting kernel density estimates (with
the bandwidth fixed from before).

The first metric we consider is the percentage of students eligible
for free school meals, an often-used widening participation criterion
in the UK which can be found in Edubase. Free school meals are a statutory
benefit available to school-aged children from families who receive
other income-assessed government benefits and can be used as a proxy
of the economic status of a school\textquoteright s students. Figure~\ref{fig:metrics}a
shows that both HiSPARC and IRIS schools feature considerably lower
free schools meals percentages than the national statistics in terms
of the location (e.g. median), scale (e.g. interquartile range), and
shape (e.g. tail heaviness) of their distributions. In contrast, the
kernel density estimate for PRiSE schools appear to tend towards higher
percentages and be somewhat broader. We perform Wilcoxon rank-sum
tests, which test whether one sample is stochastically greater than
the other (often interpreted as a difference in medians) since it
is more conservative and suffers from fewer assumptions (e.g. normality,
interval-scaling) than two-sample t-tests \citep{hollander99,gibbons11}.
These tests, however, reveal that in terms of quantiles PRiSE schools
are merely consistent with the national distribution (an absolute
$+3\%$ and relative $1.26\times$ difference in medians, $p=0.708$),
as is also evident from the boxplots. This means that PRiSE is serving
schools with considerably more disadvantaged students than both HiSPARC
($+7\%$ and $2.09\times$, $p=0.033$) and IRIS ($+8\%$ and $2.18\times$,
$p=2\times10^{-4}$) in this regard.

The second metric considered is the higher education participation
rate (Figure~\ref{fig:metrics}b), which measures how likely young
people are to go on to higher education (e.g. university) based on
where they live \citep{polar}. All the programmes considered involve
schools with students from areas with greater participation in higher
education than is representative of the entire country, but they are
largely similar to one another. This is not surprising for PRiSE given
it is limited to London, since it has been noted that young people
across London are generally more likely to access higher education
than those elsewhere in the UK \citep{polar}. However, the fact that
PRiSE's results are similar to the two national programmes is perhaps
surprising as HiSPARC lists no London schools and only $17\pm3\%$
of IRIS schools are in the Greater London area. Therefore one might
expect these two programmes to have markedly lower participation rates
in higher education than PRiSE purely due to this fact, which is not
the case. Given in-person interactions between researchers and schools
is a critical part of the PRiSE model, the geographical reach of the
Queen Mary programme will always be limited to London and thus it
is difficult to effect much change on the higher education participation
rate. The expansion of PRiSE to universities in other areas however
may help address this in future.

The third metric used is the index of multiple deprivation \citep{imd},
a UK government qualitative study of deprived areas in English local
councils by: income; employment; health deprivation and disability;
education, skills, and training; barriers to housing and services;
crime; and living environment. Here we use the index of multiple deprivation
scores (Figure~\ref{fig:metrics}c), where higher scores indicate
more deprivation. Averaged over each PRiSE school's catchment area,
these scores are considerably higher than the national statistics
(the median is $+9.40$ and $1.54\times$ larger, $p=7\times10^{-4}$).
This difference, however, disappears when using the local proxy yielding
simply a representative distribution ($+1.50$ and $1.09\times$,
$p=0.371$). In contrast, IRIS schools again clearly favour fewer
disadvantaged students than PRiSE ($-6.56$ and $0.65\times$, $p=4\times10^{-4}$)
and thus also the national statistics ( $-5.07$ and $0.71\times$,
$p=2\times10^{-6}$), whereas these differences are perhaps only marginal
for HiSPARC due to small numbers ($-6.84$ and $0.64\times$, $p=0.142$
compared to PRiSE; $-5.35$ and $0.69\times$, $p=0.288$ compared
to national statistics).

The final metric used is the percentage of ethnic minorities (Figure~\ref{fig:metrics}d)
taken from census area data. The PRiSE programme features a very broad
distribution with much greater percentages of ethnic minorities than
both the 13\% of all people from ethnic minorities across the UK \citep{census11}
and those from the areas covered by the other programmes. This is
simply due to the fact that London is the most ethnically diverse
region in the country though. The distributions for IRIS and HiSPARC
are very similar to one another and while their distributions' location
parameters are just less than the overall national statistic, they
both feature greater ethnic diversity than compared to the full distribution
across all census areas (in the case of HiSPARC this is not strictly
statistically significant at $+4\%$ absolute and $1.79\times$ relative,
$p=0.056$, likely due to small number statistics).

In terms of gender balance, a similar school/area-level approach would
not capture the known bias present in physics where only $22\%$ of
A-Level physics students are female \citep{iop}. Therefore, at kick-off
meetings the percentage of young women or girls involved in PRiSE
at each school were observed (accurate to the nearest $10\%$). The
median of these is $40\%$ though there has been considerable variation
in the gender balance amongst schools. We have worked with 11 girls'
schools (compared to just 8 boys' schools) to help address gender
balance across the programme. How this variation on a school-by-school
basis compares to each schools' A-Level cohorts or across all schools
nationally is unknown, as this data is not publicly available. IRIS,
HiSPARC and ORBYTS have not yet reported on the gender variations
in their programmes.

To summarise, PRiSE has to date engaged with a much more diverse set
of schools  with significantly more under-represented groups than
other similar schemes and reflects national statistics in most measures,
sometimes even featuring higher proportions of underserved groups.
Some work, however, needs to be done for PRiSE to be more representative
in terms of schools' categories and admissions policies, which is
currently being addressed. Future work could expand the evaluation
of participation to go beyond school/area-level metrics and investigate
individual students' characteristics, however this would require funding
to commission such research by social scientists along with the necessary
ethical approval.

\section{Retention\label{sec:Retention}}

It is natural to expect some drop-off in participation of schools
for a long-term and primarily extra-curricular programme. We employ
the branch of statistics known as survival analysis \citep{miller97}
to quantify the retention of schools throughout the PRiSE programme.
In particular we calculate both survival (the probability that a school
is still involved with the programme at a certain stage) and hazard
(the probability a school still involved at a certain stage will drop
out at the next stage) functions.

\subsection{Across interventions}

First we assess the retention of schools across the intervention stages
of the programme. This is done for academic years 2017/18 and 2018/19
only, since before this our data collecting was insufficient to track
schools' retention throughout the different stages of the programme
and in 2019/2020 the programme was disrupted by the COVID-19 pandemic
just before the comments stage with the conference having to be postponed.
In many cases in our results an assumption has been made about when
schools may have dropped out, because in \textbf{$68\pm12\%$} of
schools which drop out after the kick-off stage the teachers do not
inform us and simply no longer reply to our continuing emails (with
little difference in this percentage across the different project
stages). In these cases we assume that schools dropped out at the
earliest point the teacher became unresponsive.

\begin{figure*}
\begin{centering}
\includegraphics{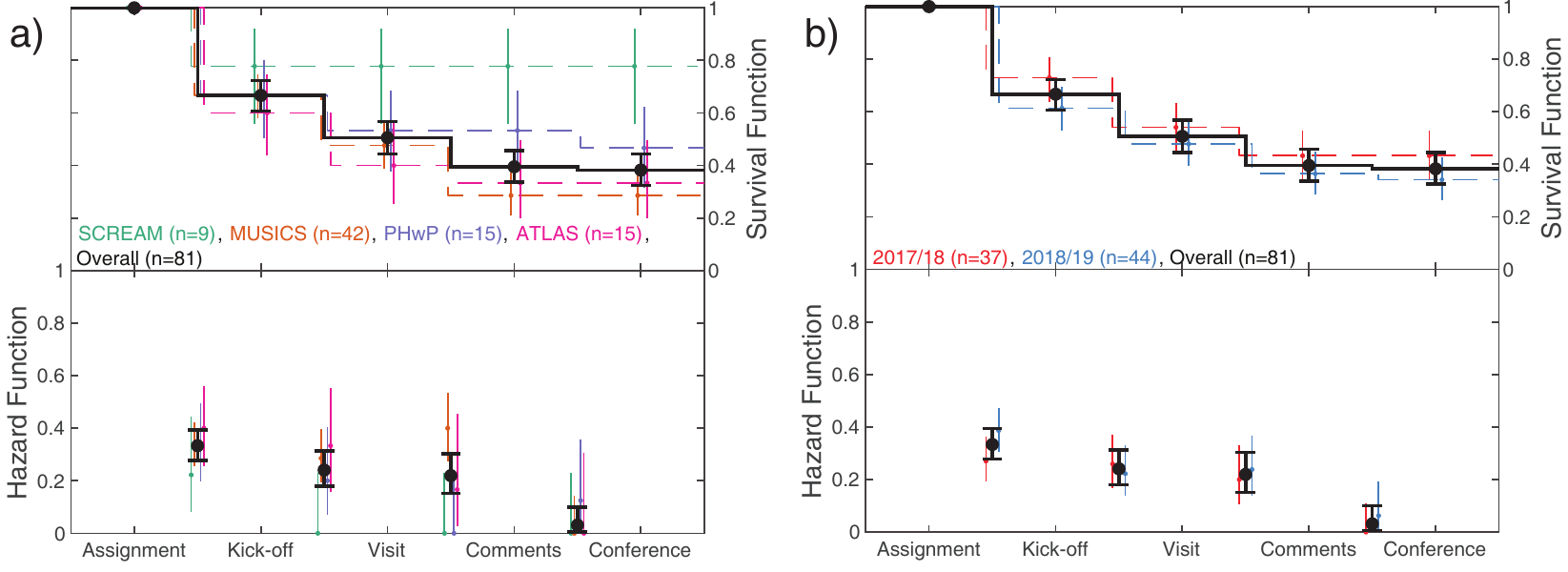}
\par\end{centering}
\caption{Schools' survival (top) and hazard (bottom) functions across interventions.
Overall results in black, colours subdivide by a) project and b) year.
Error bars denote standard \citet{clopper34} intervals.\label{fig:interventions}}
\end{figure*}

Figure~\ref{fig:interventions} shows schools retention both by project
(panel~a) and year (panel~b) across the different intervention stages.
Overall there is a fairly uniform drop-off rate between the kick-off
-- visit -- comments stages at $23\pm7\%$ with a slightly higher
drop-off from assignment to kick-off ($33\pm5\%$). Schools still
involved by the comments stage almost certainly attend the conference.
None of the other `research in schools' programmes have yet reported
on retention within their programmes and in general little research
into retention within (particularly free) programmes of multiple STEM
interventions with schools exists. However, we note that the figures
from PRiSE are at least consistent another programme --- those of
the South East Physics Network (SEPnet) Connect Physics pilot \citep{connect18}.
No overall differences across survival distributions in a logrank
test \citep{machin06} appear present by year ($\chi^{2}\left(1\right)=0.60$,
$p=0.441$) or project ($\chi^{2}\left(3\right)=4.28$, $p=0.232$).
When comparing different projects in the same year and the same projects
in different years there are differences in the survival/hazard functions
to those shown in Figure~\ref{fig:interventions} though (see data
in Appendix~\ref{sec:Retention-data}). One interpretation of the
results shown might be that projects where schools are lent equipment
(i.e. SCREAM) are more likely to succeed. Past experience across SEPnet
with the CERN@School IRIS project doesn't support this hypothesis,
however, since almost all SEPnet target schools that were loaned detectors
ended up not using them (D. Galliano, personal communication, 2018).
To prevent equipment going unused, the SCREAM project has been made
open only to schools that have successfully undertaken a different
project with us previously, which likely plays a factor in the results
presented. In general, projects based around expensive equipment are
not scalable given funding limitations so would necessarily always
have a limited reach.

While all intervention stages are offered to all schools, it should
be noted that not all schools actively engaged in project work take
advantage of them: $7\pm4\%$ don't attend a kick-off, $39\pm9\%$
do not schedule a researcher visit, and $47\pm10\%$ don't solicit
comments on their work. Schools which don't solicit a researcher visit
are much more likely to subsequently drop out ($41\pm14\%$) compared
to those which do ($12\pm8\%$). This highlights the importance of
researcher-involvement in the success of programmes, despite independent
research projects in schools often not being supported by external
mentors in general \citep{bennett16,bennett18}. The other two intervention
stages do not appear to be critical for schools' retention. Firstly,
in the case of the kick-off, this suggests the resources provided
are sufficient to still undertake project work. Secondly, while comments
on draft presentations are valued by students and teachers, students
are still able to produce work that can be presented at a conference
without them.

In addition to equality in access, it is also important that programmes
be equitable so that everyone has a fair chance of succeeding. We
find no biases in schools' ability to successfully complete a year
by school category ($\chi^{2}\left(4\right)=3.21$, $p=0.524$), free
schools meals percentage ($+4\%$ absolute and $1.33\times$ relative
differences in medians of schools which do and do not complete a year,
$p=0.865$), higher education participation rate ($+2\%$ and $1.04\times$,
$p=0.677$), indices of multiple deprivation ($-4.30$ and $0.86\times$,
$p=0.219$), or ethnic diversity ($-3\%$ and $0.93\times$, $p=0.844$).
We do not perform this test for gender variation since the data is
less reliable, as previously mentioned. Therefore, the typical societal
barriers to STEM do not appear to affect schools' ability to succeed
within the PRiSE framework.

We note that as well as schools dropping out, even within those schools
which complete a year there is typically some reduction in the number
of students that persist with project work. As our reporting only
recorded total numbers of students by Key Stage (see Table~\ref{tab:school-years}
for more information) at events rather than on a school-by-school
basis, we cannot calculate student retention rates for schools which
received on campus kick-off events, since these involved multiple
schools some of which subsequently dropped out. Neglecting those schools,
we find that between 2015--2019 the overall retention rate from kick-off
to conference was $56\pm3\%$ ($n=321)$, though we note individual
schools' rates varied widely with an interquartile range of 39--88\%
and 7~schools (out of the 27 considered) retained all students throughout
the programme. Similar to with participation, future work (subject
to funding and ethics approval) could investigate the retention within
PRiSE at the student-level and whether this is also equitable.

\subsection{Across years}

\begin{figure}
\begin{centering}
\includegraphics{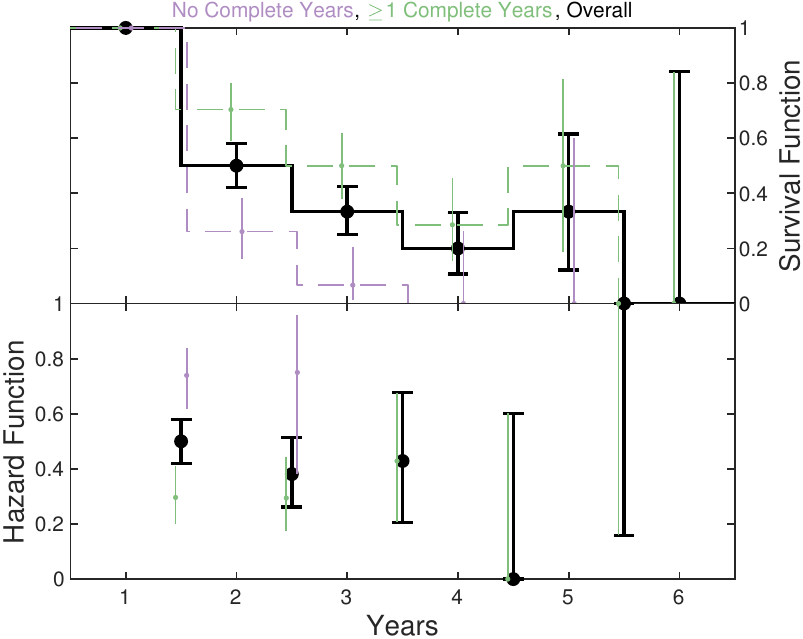}
\par\end{centering}
\caption{Schools' survival (top) and hazard (bottom) functions across years
in a similar format to Figure~\ref{fig:interventions}. Overall results
in black, with colours subdividing by schools which did (green) and
did not (purple) complete at least one year.\label{fig:years}}
\end{figure}

We have seen considerable repeated buy-in from many schools over multiple
years with PRiSE (see Appendix~\ref{sec:PRiSE-schools-table} for
the data), thus we also investigate retention across years. Figure~\ref{fig:years}
shows the overall results in black where only schools which began
work on projects are included. We note that because the programme
has been carefully grown since its inception, not all schools started
at the same time and this is why the survival function is not strictly
decreasing, e.g. only 6~schools could have been involved for 5~years
and just one for the full 6~years. Overall the drop-off rate is consistent
year-on-year at $45\pm10\%\,\mathrm{year}^{-1}$ (averaged over the
first 2~years for which we have better statistics). Similarly to
before, there is unfortunately no suitable benchmark for comparison
since no similar programmes have yet reported on university--school
relationships built over several years of running the same programme.
Again we find no significant differences in schools which participate
in PRiSE for multiple years compared to those which don't in terms
of school category ($\chi^{2}\left(4\right)=2.22$, $p=0.695$), free
schools meals percentage ($+5\%$ absolute and $1.43\times$ relative,
$p=0.327$), higher education participation rate ( $-1\%$ and $0.97\times$,
$p=0.458$), indices of multiple deprivation ($+0.75$ and $1.03\times$,
$p=0.371$) or ethnic diversity ($+12\%$ and $1.32\times$, $p=0.111$).
However, as shown in Figure~\ref{fig:years}, schools which have
been able to successfully complete at least one year (green) are far
more likely to participate again at $70\pm10\%$ compared to those
which have not (purple) at $26\pm11\%$. Importantly though, that
latter value is not negligible showing that some schools are willing
to try again. So far (not including academic year 2019/20) there have
been three out of a potential seven second attempts where this has
led to subsequent success, with these largely being run with the same
teacher rather than a different one.

\begin{figure}
\begin{centering}
\includegraphics{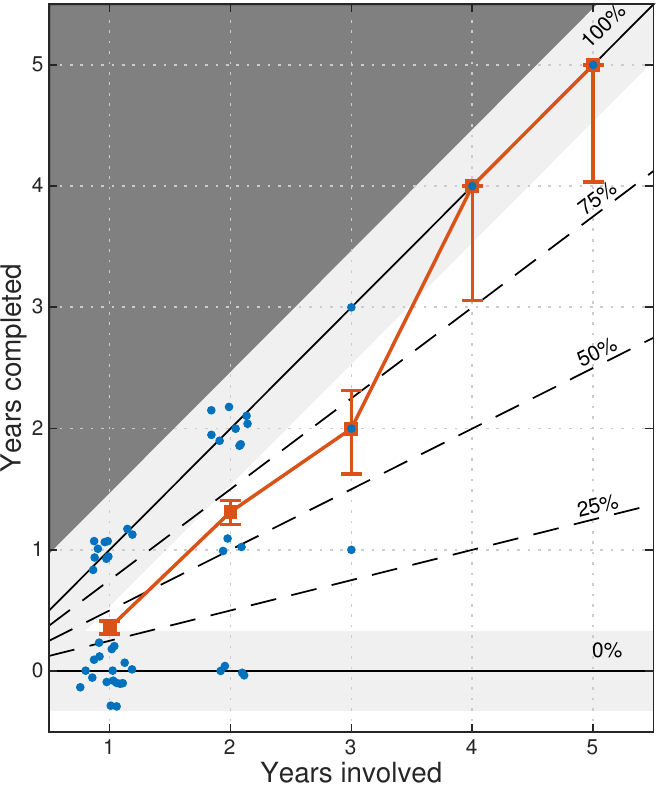}
\par\end{centering}
\caption{The number of years each school completed a PRiSE project against
the number of years they were involved. Datapoints (blue) have been
jittered for visibility. The average number of years completed and
standard \citet{clopper34} interval for this rate are also shown
(red).\label{fig:years-vs-complete}}

\end{figure}

The link between the success of schools at completing projects and
the continued involvement of schools across several years is highlighted
further in Figure~\ref{fig:years-vs-complete}. This shows for each
school listed in Appendix~\ref{sec:PRiSE-schools-table} the number
of years that they completed PRiSE projects against the total number
of years they started them. Note that we do not include academic year
2019/20 in either dimension due to disruption by the COVID-19 pandemic.
It appears that schools predominantly tend to lie near the two possible
extremes (solid lines) of either completing projects every year they
were involved or not completing the projects at all. The average rate
and its uncertainty (constructed by treating each school as a single
binomial experiment and combining them using Bayes' theorem) are also
shown in the figure in red as a function of the number of years involved.
A clear trend can be seen that completion rates increased as schools
were involved for more years. However, one must be careful in interpreting
possible causation behind this correlation. While there are examples
where a second attempt at a PRiSE project did lead to improved retention,
in most cases it is likely that schools which succeed are simply more
likely to participate again whereas those that (for whatever reason)
don't succeed are understandably less likely to continue in subsequent
years.

While at our conferences teachers are unanimous in their intentions
to participate again when asked via paper questionnaire, the data
in Figure~\ref{fig:years} show a fairly consistent drop-off rate
of $30\pm12\%\,\mathrm{year}^{-1}$ from schools which completed a
year. The majority of this might be explained by teacher turnover,
which in 2015 stood at $21\%$ per year nationally ($13\%$ left teaching
altogether) with an increasing year-on-year trend and London having
been highlighted as having a much higher churn rate \citep{worth}.
We are aware (often through out of office messages or bounced emails)
of numerous teachers having either moved schools or left the profession
and it is rare that they communicate this with us ahead of time, handover
project responsibilities to another teacher, or take projects with
them to their new school. Establishing the exact number of schools
which did not return due to this reason compared to other explanations
would require considerable resources beyond the scope of this evaluation
though. Teacher turnover poses a real challenge to establishing lasting
relationships with schools, when these relationships are so dependent
on individuals and are often not embedded within the schools themselves
--- though the same may also be said of universities, e.g. most of
the public engagement professionals who led the Beacons programmes
in the UK \citep{nccpebeacons} don't work in public engagement anymore.

Maximising the retention of schools across years is necessarily a
function of the capacity of a programme. While bringing new schools
into the programme is certainly beneficial, we have seen that teachers'
ability and confidence in supporting project work in their school
develops not only across the 6-month programme but over several years.
Therefore, a practical balance could be to aim to involve schools
directly for just a few years to the point that they can sustainably
run projects with fewer interventions from the university, perhaps
just an on campus kick-off and then the conference as researcher capacity
is less of an issue with these. We acknowledge though that some schools
might no longer participate without the full suite of interventions.
This approach would enable the wider impacts on schools that benefit
from multiple years of participation while also ultimately freeing
up capacity in the long-term for new schools to be able to benefit
from the programme.

\section{Feedback\label{sec:Feedback}}

Here we present qualitative data from teachers relating to the issues
of accessibility, diversity, equity, and retention.

\subsection{Method}

We prefer to take a holistic approach in investigating teachers' thoughts
about these issues, using evaluative data from a variety of methods
throughout the programme. Formal feedback from teachers has been gathered
from teachers via paper questionnaires handed out at our student conferences
each year (in 2020 due to the COVID-19 pandemic an online form was
instead used). All data used here were in response to open questions
and further details of this questionnaire can be found in \citet{archer_prise_framework20}.
In addition to this formal feedback, we also use data obtained more
informally. These include comments made by teachers in-person throughout
the programme, e.g. during researcher visits or our conference, and
those passed on via email. Where possible explicit consent has been
obtained to use these comments, though in general by participating
in PRiSE teachers are aware that they are entering a relationship
of mutual trust and that any information passed on by them will be
used with integrity \citep[cf.][]{bera18}. These informal comments
are recorded at the time and then analysed along with and in the same
way as the formal feedback. The anonymity of the teachers and their
schools are protected, quoting only the school's pseudonym as well
as the project and year the feedback related to along with the method
the data was obtained.

All qualitative data were analysed using thematic analysis \citep{Braun2006},
with the themes being allowed to emerge from the data via grounded
theory \citep{Robson2011,Silverman2010} as follows:
\begin{enumerate}
\item Familiarisation: Responses are read and initial thoughts noted.
\item Induction: Initial codes are generated based on review of the data.
\item Thematic Review: Codes are used to generate themes and identify associated
data.
\item Application: Codes are reviewed through application to the full data
set.
\item Analysis: Thematic overview of the data is confirmed, with examples
chosen from the data to illustrate the themes.
\end{enumerate}
In the following subsection, we highlight the different themes identified
relating to aspects of accessibility, diversity, equity, and retention
through bold text, providing illustrative quotes.

\subsection{Results and discussion}

Several teachers from a variety of different schools have raised that
they \textbf{value the diversity} present across PRiSE, particularly
that students from different schools and backgrounds are able to interact
as equals at the student conferences.\begin{quote}``\textit{They
love the competition with independent schools}'' (Teacher, Rushmore
Academy, MUSICS 2018, in-person)\\
``\textit{Giving the students the opportunity to meet with other
schools and academic staff in person is a major highlight}'' (Teacher,
Tree Hill High School, PHwP 2020, questionnaire)\end{quote}One teacher
expanded on this, contrasting the diversity present to other schemes
and positing that the support provided through PRiSE enabled this
difference\begin{quote}``\textit{At }{[}PRiSE student conference{]}\textit{
Cosmic Con, pupils from a diverse range of state and independent schools
have the opportunity to share their experiences and discuss their
findings with each other, widening their perspective from the natural
micro-habitat of the school environment to the wider community around
them. Groups that have worked on the same project naturally gravitate
into discussion. The key to increasing collaboration and empathy for
those in different schools is to put pupils in the same place, at
the same time, with some interest or experience in common to talk
about. There are other schools conferences around, but QMUL has such
close links with and provides such support for schools, that this
adds to the diversity of Cosmic Con.}'' (Teacher, Octavian Country
Day School, PHwP 2019, email testimonial following in-person comment)\end{quote}The
equality of all students at PRiSE conferences is further reflected
in that prizes awarded to students, judged by researchers, have been
well distributed amongst different types of schools with no obvious
biases (though due to the small numbers we do not perform a thorough
statistical analysis). This has, however, had a prodigious effect
on students from disadvantaged backgrounds\begin{quote}``\textit{The
students were buzzing on their journey back, they kept saying `I can't
believe our little comprehensive won'.}'' (Teacher, Coal Hill School,
MUSICS 2016, in-person)\end{quote}

Teachers' comments have highlighted that while some PRiSE projects
are thought to be equitable in terms of \textbf{students' ability},
that unfortunately may not currently the case with all of them\begin{quote}``\textit{The
initial information received on the project was quite daunting. But
the presentation introduction to it was much more accessible. The
scope of the project was accessible to students of all abilities.}''
(Teacher, Spence Academy for Young Ladies, MUSICS 2018, questionnaire)\\
``\textit{It is really appropriate in content and context for my
students... Throughout the project there have been tasks that can
meet the varying abilities of my students. Really well designed project.}''
(Teacher, Bending State College, PHwP 2020, questionnaire)\\
``\textit{It gives students a chance to do something very interesting,
if they are enthusiastic. There isn't much to do for students who
are struggling. The projects are very high demand, both in time and
skill. This has always been our problem with retaining students' interest
over several months. I'm not sure how it can be changed too much without
making it boring for students, it's a difficult balance. I think just
a variety in skill level, it's nice to have a whole class working
on a project together because it pulls everyone up, so there should
be stuff for the E/D-grade students to have a go at as well as the
A/A{*} pupils. Some projects don't lend themselves very well to that,
but others like MUSICS do because anyone can listen to some sounds
but there was still stuff for the higher skill students to get stuck
into.}'' (Teacher, Hill Valley High School, ATLAS 2020, questionnaire)\end{quote}In
addition to simply tweaking the projects to reduce their barrier to
entry while still maintaining their broad open-ended scope, there
are other potential ways to address this also. One teacher (Xavier's
Institute for Higher Learning, MUSICS 2017) highlighted in-person
that the students' group dynamics can play a big role in their successful
participation. In the previous year a clear enthusiastic leader had
emerged who could include everyone in the project in different ways
appropriate to their ability, whereas such leadership had not successfully
been established amongst any of the students the following year leading
to the group struggling to find a direction and cultivate eagerness
in everyone participating until the researcher visit occurred. We
suggest that teachers might be able to help facilitate establishing
the group dynamics where required, since they are more familiar with
the students. Another possibility is peer mentoring from the school's
previous year of PRiSE students.\begin{quote}``\textit{I have had
a lot of help from Year~13 students acting as mentors... this has
been a useful exercise in peer learning.}'' (Teacher, Xavier's Institute
for Higher Learning, MUSICS 2017, questionnaire)\end{quote}Of course
this is only a viable option for schools which complete one year in
the first instance. Some teachers who have expressed interest in capitalising
on peer mentoring have often struggled to implement it within their
schools though. Furthermore, some schools prefer to change project
after a couple of years and so the ability of previous PRiSE students
to effectively mentor, apart from in a more pastoral capacity, may
be limited. While further research is required into what makes for
successful group dynamics or student mentorship, we nonetheless hope
to be able to include something on both of these aspects in planned
`how to' guides for teachers.

Finally, we discuss the issue of retention within the programme. Unfortunately,
given the often poor \textbf{communication} from teachers (also highlighted
as an issue within the ORBYTS programme, \citealp{sousasilva17})
it is not always clear as to why individual schools drop out. We have
little evidence around why teachers do not communicate this, though
one teacher (Colonial Fleet Academy, MUSICS 2016, email) who had been
unresponsive eventually expressed a feeling of embarrassment that
all their students bar one had ceased project work. While we try and
assure teachers upon initiating the projects that we anticipate some
drop-off and that there is no pressure from us to remain involved,
more may be required in this area. From the minority of teachers that
do inform us of their school dropping out, typically via email, reasons
have included mock exams getting in the way, students losing interest,
realising the amount of work involved, difficulty balancing the project
with other activities / their normal school workload, giving up due
to uncertainty in how to progress, and not feeling like they've made
enough progress to continue. Similar themes have been expressed from
teachers in-person concerning some (but not all) of their students
dropping out. We feel that many of these issues might have been mitigated
through more and/or earlier communication from the school to us, as
we have been able to assist with similar struggles at more communicative
schools. Despite a clear support process being laid out at kick-off
workshops a number of teachers from schools that successfully completed
projects realised in hindsight that they should have taken advantage
of the opportunities from the university earlier in the programme
than they did\begin{quote}``\textit{Should have taken advantage
}{[}of support from Queen Mary{]}\textit{ at }{[}an{]}\textit{ earlier
stage.}'' (Teacher, St Trinians, SCREAM 2018, questionnaire)\\
``\textit{The call with Dr Senz would have been more helpful earlier
on - it made a big difference, and students would have got more out
of the project if they had more time after this took place.}'' (Teacher,
Imperial Academy, ATLAS 2020, questionnaire)\end{quote}This is again
something that could be stressed in teacher guidance upon engaging
with the programme to better set expectations and good practice, as
many new teachers to the programme may not be used to such a reactive
way of working rather than the typical `push' model from teachers
to students.

Overall, we get the impression that retention within PRiSE can often
come down to the individual teacher -- those that are communicative
and properly engage with the programme and its expectations from the
outset are far more likely to see their students succeed. This is
backed up by evidence from several schools, where a change of teacher
has either led to increased engagement with the programme (e.g. Rydell
High School), success at previously unsuccessful schools (e.g. Hill
Valley High School), or unfortunately previously successful schools
dropping out of the programme (e.g. Prufrock Preparatory School).
One teacher who changed schools (from Hogwarts to Prufrock Preparatory
School) bringing PRiSE projects with them also raised with us in-person
that the culture within the schools can play a role in students' engagement
with the programme (and extra-curricular activities in general). Both
of these are challenging issues to address as we aim to increase the
retention and equity of the programme. While it is clear that more
detailed qualitative research is required in this area, perhaps giving
clearer information and expectations upon signup as well as providing
further guidance on ways of successfully integrating and nurturing
project work within schools, as highlighted by other teachers, these
issues might be somewhat mitigated.

\conclusions{}

Societal inequalities in access to and engagement with science are
prevalent even in secondary/high schools \citep{wellcome19}. While
university engagement programmes, like independent research projects,
could address these issues, at present few such schemes specifically
target traditionally under-represented groups and in general globally
students' participation in such projects are inequitable \citep{bennett16,bennett18}.
In this paper we have evaluated the accessibility, diversity, and
equity of the `Physics Research in School Environments' (PRiSE) programme
of independent research projects \citep{archer_prise_framework20}.

The schools involved in PRiSE have been benchmarked against those
participating in similar programmes of research-based physics projects
for schools in the UK. Investigating measures of the socio-economic
status, race, and genders of the schools' students have revealed that
PRiSE has engaged much more diverse groups of schools  with substantially
more under-represented groups than is typical. Indeed, PRiSE schools
are mostly reflective of national statistics and in some measures
feature an over-representation of disadvantaged groups. While PRiSE
has featured fewer independent and selective schools than other schemes,
the proportions are not currently reflective of all schools nationally
and thus new policies have been implemented to improve diversity in
these regards.

Survival analysis has been used to explore the retention of schools
within the programme. This was firstly done across the different intervention
stages of PRiSE within each academic year. We find a fairly consistent
drop-off rate throughout, with no significant differences between
the different projects or the years considered. While little research
into the retention of schools within protracted programmes of engagement
currently exist, the rates exhibited by PRiSE are at least similar
to another programme \citep{connect18}. The analysis has highlighted
the importance of PRiSE's researcher-involvement in the schools' success.
This is despite independent research projects in general often not
being supported by external mentors \citep{bennett16,bennett18}.
No biases in schools' retention appear present by school category,
socio-economic background, or race. This suggests that schools'  ability
to succeed at independent research projects is independent of background
within the PRiSE framework. PRiSE has seen repeated buy-in over multiple
years from numerous schools. Hence we also looked at the retention
of schools across multiple years, again finding no real differences
in the backgrounds of schools which return and those which do not.
Indeed, the only predictor for multiple years of participation is
whether the school engaged with the programme through to completion
for at least one year. Our interpretation is that success within PRiSE
often comes down to the individual teacher, with poor communication
\citep[cf.][]{sousasilva17} or not fully engaging with the programme
and its expectations serving as key barriers in schools' participation.

Qualitative feedback from teachers have shown that they value the
diversity within the programme, seeing the ability of students from
different schools and backgrounds to interact as equals at PRiSE conferences
as a positive aspect. They also attribute this equity to the exceptionally
high level of support provided by PRiSE to the schools involved. The
need for slight modifications to make some of the projects more accessible
to students of all abilities has been raised. These concerns might
also be addressed by prompting teachers to facilitate students' group
dynamics and potentially incorporating peer mentoring from previous
years' PRiSE students, both of which have been reported as successful
in some cases but with mixed results at other schools. More teacher
guidance, co-created with teachers themselves, on the expectations
within the programme as well as good practice in incorporating and
nurturing project work within schools could be provided to help with
retention in new schools.

Our analysis has been limited to the London geographic area, so it
is not yet clear that the PRiSE framework of independent research
projects would necessarily be as accessible or equitable in different
parts of the UK or in different countries. With the adoption of this
approach to engagement at other universities, however, we hope to
be able to investigate this in the future. Further, only school-level
metrics have been considered here and more detailed analysis at the
individual student level and their characteristics could be considered
in future, which would require funding to commission such research
by social scientists along with the necessary ethical approval. Finally,
in-depth qualitative research into the reasons behind schools dropping
out of PRiSE, both within the six-month programme and between different
years, would be beneficial in understanding what the current barriers
to prolonged participation are and how these could be best addressed
in the future.

\appendix

\section{PRiSE schools\label{sec:PRiSE-schools-table}}

Below is a table with information about all the schools which have
been involved in PRiSE. To protect the anonymity of students and teachers,
pseudonyms (taken from https://annex.fandom.com/wiki/List\_of\_fictional\_schools)
have been used. The table details the year the school joined, how
many years they have undertaken projects, the number of these years
they successfully completed (i.e. made it all the way to the student
conference), along with categorical information, and deciles (used
here to further protect anonymity) across their catchment areas of
the schools' percentage of students on free school meals (FSM), higher
education participation rate (HEPR), indices of multiple deprivation
score (IMD), and percentage of ethnic minorities (EM). Further information
about these are given in Appendices~\ref{sec:school-types} and \ref{sec:metrics-methods}.
Missing data in the table is due to it not being publicly available.
Different schools partnerships where schools have worked together
(or have at least attempted to) are indicated by letters. Schools
which signed up for but never commenced project work (by hosting/attending
a kick-off meeting) are not included here. Note that the years completed
column does not include data from the 2019/20 academic year due to
disruption by the COVID-19 pandemic and we mark all schools affected
by this with an asterisk ({*}) in this column.

\begin{tiny}%
\begin{tabular}{clcccccccccc}
\textbf{Joined} & \textbf{School pseudonym} & \textbf{Category} & \textbf{Admissions} & \textbf{Gender} & \textbf{FSM} & \textbf{HEPR} & \textbf{IMD} & \textbf{EM} & \textbf{Years} & \textbf{Completed} & \textbf{Partnership}\tabularnewline
\hline 
2014 & Hogwarts & Independent &  & Mixed &  &  &  &  & 5 & 5 & a\tabularnewline
2015 & Colonial Fleet Academy & Academy & Non-Selective & Mixed & 7 & 2 & 8 & 8 & 1 & 1 & \tabularnewline
2015 & Constance Billard School for Girls & LA maintained & Non-Selective & Girls & 8 & 6 & 9 & 10 & 2 & 0 & b\tabularnewline
2015 & St. Judes School for Boys & LA maintained & Non-Selective & Boys & 9 & 6 & 9 & 10 & 2 & 0 & b\tabularnewline
2015 & Sweet Valley High School & Academy & Non-Selective & Mixed & 5 & 9 & 7 & 10 & 2 & 2 & \tabularnewline
2015 & Xavier's Institute for Higher Learning & College &  & Mixed &   &  &  &  & 5 & 4{*} & \tabularnewline
2016 & Angel Grove High School & LA maintained & Non-Selective & Girls & 6 & 7 & 7 & 10 & 1 & 1 & c\tabularnewline
2016 & Coal Hill School & LA maintained & Non-Selective & Girls & 6 & 6 & 7 & 9 & 2 & 2 & \tabularnewline
2016 & Earth Force Academy & Academy & Non-Selective & Mixed & 3 & 8 & 4 & 8 & 1 & 1 & \tabularnewline
2016 & Hill Valley High School & LA maintained & Non-Selective & Mixed & 8 & 6 & 9 & 10 & 3 & 1{*} & d\tabularnewline
2016 & Hillside Academy & Academy &  & Mixed &   &  &  &  & 1 & 0 & \tabularnewline
2016 & Imperial Academy & Academy & Selective & Boys & 1 & 9 & 4 & 9 & 3 & 2{*} & e\tabularnewline
2016 & Our Lady of Perpetual Sorrow & Independent &  & Girls &   &  &  &  & 1 & 0 & \tabularnewline
2016 & Prufrock Preparatory School & Independent &  & Mixed &   &  &  &  & 3 & 2 & a\tabularnewline
2016 & Springfield Community College & College &  & Mixed &   &  &  &  & 1 & 1 & \tabularnewline
2016 & St Trinians & Independent &  & Girls &   &  &  &  & 4 & 3{*} & a\tabularnewline
2016 & Stone Canyon High School & LA maintained & Non-Selective & Boys & 7 & 7 & 8 & 10 & 1 & 1 & c\tabularnewline
2016 & Stoneybrook Academy & Academy &  & Mixed &   &  &  &  & 2 & 0 & \tabularnewline
2016 & Tree Hill High School & LA maintained & Non-Selective & Mixed & 7 & 9 & 6 & 10 & 4 & 1{*} & d\tabularnewline
2016 & Vulcan Science Academy & Academy & Non-Selective & Mixed & 6 & 10 & 6 & 9 & 1 & 0 & a\tabularnewline
2017 & Avalanche Arts Academy & Academy & Non-Selective & Mixed & 3 &  &  &  & 1 & 0 & \tabularnewline
2017 & Barcliff Academy & Academy & Non-Selective & Mixed & 5 & 8 & 8 & 10 & 1 & 0 & \tabularnewline
2017 & Boston Bay College & College &  & Mixed &   &  &  &  & 3 & 2{*} & \tabularnewline
2017 & Bronto Crane Academy & Academy & Non-Selective & Mixed & 8 & 8 & 7 & 10 & 2 & 1 & d\tabularnewline
2017 & Chalet School & Independent &  & Boys &   &  &  &  & 1 & 0 & \tabularnewline
2017 & Fire Nation Academy for Girls & LA maintained & Non-Selective & Girls & 8 & 8 & 8 & 9 & 1 & 0 & f\tabularnewline
2017 & Harbor School & Independent &  & Mixed &   &  &  &  & 3 & 2{*} & f\tabularnewline
2017 & Io House & Independent &  & Mixed &   &  &  &  & 1 & 1 & \tabularnewline
2017 & Kelsey Grammar School & Academy & Selective & Mixed & 1 &  &  &  & 1 & 0 & \tabularnewline
2017 & Martha Graham Academy & Academy & Non-Selective & Mixed & 5 & 8 & 8 & 9 & 3 & 0{*} & f\tabularnewline
2017 & Miss Shannon's School for Girls & LA maintained & Non-Selective & Girls & 4 & 8 & 6 & 9 & 1 & 0 & \tabularnewline
2017 & Roosevelt High & LA maintained & Non-Selective & Mixed & 8 & 4 & 9 & 9 & 2 & 1 & \tabularnewline
2017 & Rydell High School & Free School & Non-Selective & Mixed & 1 & 10 & 5 & 10 & 3 & 2{*} & \tabularnewline
2017 & Smeltings & Independent &  & Boys &   &  &  &  & 2 & 2 & \tabularnewline
2017 & Spence Academy for Young Ladies & Academy & Selective & Girls & 1 & 8 & 4 & 8 & 3 & 2{*} & \tabularnewline
2017 & St. Francis Academy High School & Academy & Non-Selective & Mixed & 7 & 10 & 5 & 9 & 1 & 0 & \tabularnewline
2017 & Stoneybrook Day School & Independent &  & Mixed &   &  &  &  & 1 & 1 & \tabularnewline
2017 & Sunnydale High School & Academy & Non-Selective & Mixed & 9 & 8 & 9 & 9 & 3 & 2{*} & \tabularnewline
2017 & Washington Preparatory Academy & Independent &  & Mixed &   &  &  &  & 1 & 0 & \tabularnewline
2018 & American Eagle Christian School & LA maintained & Non-Selective & Mixed & 3 & 7 & 7 & 10 & 1 & 0 & \tabularnewline
2018 & Bel-Air Academy & Academy & Non-Selective & Mixed & 7 & 8 & 8 & 10 & 2 & 0{*} & \tabularnewline
2018 & Bending State College & College &  & Mixed &   &  &  &  & 1 & 0{*} & \tabularnewline
2018 & Octavian Country Day School & Independent &  & Mixed &   &  &  &  & 2 & 1{*} & \tabularnewline
2018 & Pok\'{e}mon Technical Institute & LA maintained & Non-Selective & Mixed & 9 & 7 & 9 & 10 & 2 & 1{*} & \tabularnewline
2018 & Prescott Academy for the Gifted & LA maintained & Selective & Boys & 3 & 9 & 6 & 10 & 1 & 1 & \tabularnewline
2018 & Rushmore Academy & Academy & Non-Selective & Mixed & 8 & 10 & 6 & 9 & 1 & 0 & \tabularnewline
2018 & Summer Heights High & LA maintained & Non-Selective & Mixed & 5 & 7 & 8 & 10 & 1 & 0 & \tabularnewline
2018 & Thomas Aquinas Private Girls' School & Independent &  & Girls &   &  &  &  & 1 & 0 & \tabularnewline
2018 & Walford High School & Free School &  & Mixed & 1 &  &  &  & 2 & 0{*} & \tabularnewline
2018 & Worcestershire Academy & Academy & Non-Selective & Mixed & 8 & 7 & 9 & 10 & 1 & 0 & \tabularnewline
2019 & Bullworth Academy & Academy & Non-Selective & Mixed & 10 & 8 & 9 & 10 & 1 & {*} & \tabularnewline
2019 & Chuck Norris Grammar School & Academy & Non-Selective & Boys & 5 & 7 & 8 & 9 & 1 & {*} & \tabularnewline
2019 & Holy Forest Academy & Academy & Non-Selective & Mixed & 7 & 6 & 7 & 10 & 1 & {*} & \tabularnewline
2019 & Jedi Academy & Academy & Selective & Girls & 2 & 8 & 5 & 9 & 1 & {*} & e\tabularnewline
2019 & Marlin Academy & Academy & Non-Selective & Boys & 6 &  &  &  & 1 & {*} & \tabularnewline
2019 & Morningwood Academy & Academy & Non-Selective & Mixed & 2 & 7 & 4 & 9 & 1 & {*} & \tabularnewline
2019 & Orbit High School & Free School &  & Mixed & 1 &  &  &  & 1 & {*} & g\tabularnewline
2019 & Quirm College for Young Ladies & Independent &  & Girls &   &  &  &  & 1 & {*} & h\tabularnewline
2019 & Royal Dominion College & Independent & Selective & Mixed &   &  &  &  & 1 & {*} & \tabularnewline
2019 & Sky High & Free School &  & Mixed & 5 &  &  &  & 1 & {*} & h\tabularnewline
2019 & Smallville High School & Free School & Selective & Mixed &   &  &  &  & 1 & {*} & \tabularnewline
2019 & Starfleet Academy & Academy & Non-Selective & Mixed & 9 & 8 & 9 & 10 & 1 & {*} & \tabularnewline
2019 & Stoolbend High School & Special school &  & Mixed & 10 &  &  &  & 1 & {*} & \tabularnewline
2019 & Summer Bay High & LA maintained & Non-Selective & Mixed & 7 & 8 & 8 & 10 & 1 & {*} & \tabularnewline
2019 & Sycamore Secondary School & Free School & Non-Selective & Mixed & 8 & 3 & 8 & 9 & 1 & {*} & h\tabularnewline
2019 & Warren Greeley Preparatory School & Independent &  & Mixed &   &  &  &  & 1 & {*} & g\tabularnewline
2019 & Welton Academy & Academy & Non-Selective & Mixed & 4 & 10 & 3 & 8 & 1 & {*} & h\tabularnewline
\end{tabular}\end{tiny}

\section{Information about UK/English schools\label{sec:school-types}}

This paper uses the context of the UK/English education system to
assess the diversity of schools engaged within the PRiSE (and other)
programme(s). To those unfamiliar with this system, we provide some
further notes here. Schools are classified by the \citet{edubase}
into the following main categories:
\begin{itemize}
\item \textbf{Academies} are schools that are state funded and free to students
but are not run by the local authority. They have much more independence
than most other schools including the power to direct their own curriculum.
Academies are established by sponsors from business, faith or voluntary
groups in partnership with the Department for Education working with
the community.
\item \textbf{Colleges} are post-16 education establishments not part of
a secondary school.
\item \textbf{Free Schools} are a type of academy set up by teachers, parents,
existing schools, educational charities, universities, or community
groups.
\item \textbf{Independent Schools} are funded by the fees paid by the parents
of pupils, contributions from supporting bodies and investments. They
are not funded or run by central government or a Local Authority.
They can set their own curriculum.
\item \textbf{Local Authority (LA) Maintained Schools} are wholly owned
and maintained by Local Authorities and follow the national curriculum.
\end{itemize}
Other less common categories are not considered in this paper due
to small number statistics. In addition to the school category, UK
schools can also be classified by their admissions policy. Selective
(or grammar) schools enrol pupils based on ability whereas non-selective
(or comprehensive) schools are not able to do this. While some independent
schools are selective, not all of them are. Table~\ref{tab:school-years}
shows how school years are denoted in the English system, with other
contextual information.

\begin{table}
\begin{centering}
\begin{tabular}{|c|c|c|c|c|}
\hline 
\textbf{Key Stage} & \textbf{Year} & \textbf{Final Exam} & \textbf{Age} & \textbf{Policy}\tabularnewline
\hline 
\hline 
\multirow{2}{*}{KS4} & 10 & None & 14--15 & \multirow{2}{*}{Compulsory}\tabularnewline
\cline{2-4} \cline{3-4} \cline{4-4} 
 & 11 & GCSE & 15--16 & \tabularnewline
\hline 
\multirow{2}{*}{KS5} & 12 & AS-Level (optional) & 16--17 & \multirow{2}{*}{Optional}\tabularnewline
\cline{2-4} \cline{3-4} \cline{4-4} 
 & 13 & A-Level & 17--18 & \tabularnewline
\hline 
\end{tabular}
\par\end{centering}
\caption{Summary of the stages in the English education system applicable to
PRiSE.\label{tab:school-years}}

\end{table}

\section{Method for gathering metrics on UK schools\label{sec:metrics-methods}}

In this paper we look at several metrics to assess the backgrounds
of the schools' pupils. School categories are listed in Edubase \citep{edubase},
as are the percentage of students on free school meals (though this
is typically not listed for colleges, independent schools, and some
academies). Other relevant metrics are not included in Edubase but
are tied to census lower and middle layer super output areas (LSOAs
and MSOAs respectively, \citealp{census11}). However, since schools
will draw students from a wider range of locations than simply the
census area within which they are located, this ideally necessitates
knowledge of a school's catchment area to gain a better understanding
of the backgrounds of the schools' students. While such information
is available for state schools in the Greater London area through
the London Schools Atlas \citep{londonschoolsatlas}, which lists
for each school the resident LSOAs/MSOAs of their student base, unfortunately
there is no equivalent publicly accessible data covering the entire
UK.

Here we detail how school metrics across their entire catchment areas
are calculated. Higher education participation rates for each school
are determined using the number of entrants to higher education and
cohort populations across each school's MSOAs from POLAR4 data \citep{polar}.
Similarly, index of multiple deprivation scores \citep{imd} are averaged
over each school's LSOAs. For protected characteristics such as gender/sex
and race/ethnicity, we do not collect this data from students for
ethics reasons. In the latter case, while it has been observed by
session leaders at interventions that a diversity of ethnicities have
been involved, we opt to quantify this through the ethnic diversity
of the areas from which students are drawn. Census data \citep{census11}
on ethnic groups is used to calculate the percentage of people from
ethnic minorities (i.e. non-white groups) across each school's LSOAs.

\begin{figure*}
\begin{centering}
\includegraphics{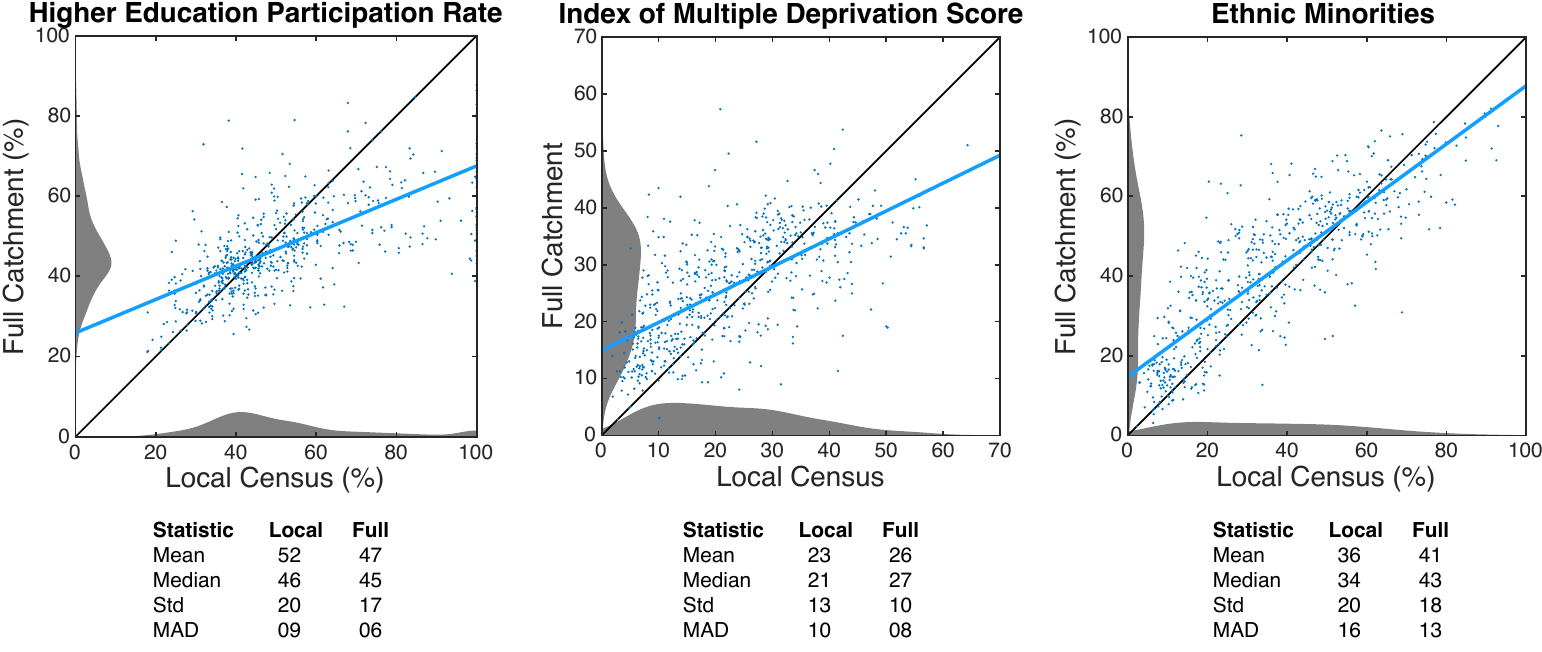}
\par\end{centering}
\caption{Comparison of local census and full catchment data for all schools
across Greater London ($n=625$), with a linear regression (blue line)
and marginal distributions (grey areas) shown. Location (mean and
median) and scale (standard deviation and median absolute deviation)
parameters for the two types of data are also displayed.\label{fig:local-catchment}}

\end{figure*}

To still enable some comparison between PRiSE and the national programmes,
for which we do not have access to information on schools' catchment
areas, we rely on using the metrics pertaining only to the LSOA/MSOA
within which the school resides. These results are indicated by dashed
lines in Figure\textbf{~}\ref{fig:metrics} and have also been computed
for PRiSE schools to ensure like-for-like comparisons. Across the
Greater London area we can check the reliability of these local proxies
and comparisons are shown in Figure~\ref{fig:local-catchment}. These
reveal that the local and full catchment distributions appear similar,
with location and scale parameters (e.g. means and standard deviations
respectively) that differ only by a few percent / score points. Therefore,
taking into account schools' full catchment area only slightly changes
the underlying distributions of the societal measures, apart from
in the extremes of the distributions (i.e. the tails) where greater
differences occur. However, while the local census and full catchment
data for London schools correlate, this correlation is not particularly
strong (the correlation coefficients are $R=0.64$, $0.65$, and $0.85$
respectively) and the linear best fit lines have slopes significantly
less than unity. This highlights that the metrics vary substantially
across all the census areas a school draws students from, meaning
that a school can have rather different values when either using local
or full catchment data. Further investigation of these societal measures
applied to schools in general is beyond the scope of this paper.

\section{Retention data\label{sec:Retention-data}}

The below table contains data on the retention of schools within PRiSE
across intervention stages. We use the following terminology:
\begin{itemize}
\item \textbf{Attended}: schools which received the intervention
\item \textbf{Didn't attend}: schools which did not receive the intervention
but were still engaged with the programme at that stage
\item \textbf{Unresponsive}: schools that did not respond to our communications
from that point on
\item \textbf{Dropped out}: schools that communicated their dropping out
the programme at that stage
\end{itemize}
Schools that have become unresponsive or drop out are no longer counted
in the table for subsequent intervention stages.
\begin{center}
\begin{footnotesize}%
\begin{tabular}{l|l|cccc|cccc|}
\multicolumn{1}{l}{} &  & \multicolumn{4}{c|}{2017/18} & \multicolumn{4}{c|}{2018/19}\tabularnewline
\cline{3-10} \cline{4-10} \cline{5-10} \cline{6-10} \cline{7-10} \cline{8-10} \cline{9-10} \cline{10-10} 
\multicolumn{1}{l}{} &  & \begin{turn}{90}
SCREAM
\end{turn} & \begin{turn}{90}
MUSICS
\end{turn} & \begin{turn}{90}
PHwP
\end{turn} & \begin{turn}{90}
ATLAS
\end{turn} & \begin{turn}{90}
SCREAM
\end{turn} & \begin{turn}{90}
MUSICS
\end{turn} & \begin{turn}{90}
PHwP
\end{turn} & \begin{turn}{90}
ATLAS
\end{turn}\tabularnewline
\hline 
Assignment & Assigned & 5 & 18 & 7 & 7 & 4 & 24 & 8 & 8\tabularnewline
\hline 
\multirow{4}{*}{Kick-off} & Attended & 3 & 14 & 4 & 4 & 4 & 13 & 5 & 3\tabularnewline
 & Didn't attend & 0 & 0 & 1 & 1 & 0 & 1 & 0 & 1\tabularnewline
 & Unresponsive & 1 & 1 & 0 & 2 & 0 & 6 & 3 & 3\tabularnewline
 & Dropped out & 1 & 2 & 2 & 0 & 0 & 4 & 0 & 1\tabularnewline
\hline 
\multirow{4}{*}{Visit} & Attended & 3 & 7 & 1 & 1 & 4 & 5 & 4 & 0\tabularnewline
 & Didn't attend & 0 & 4 & 2 & 2 & 0 & 4 & 1 & 3\tabularnewline
 & Unresponsive & 0 & 2 & 1 & 1 & 0 & 5 & 0 & 1\tabularnewline
 & Dropped out & 0 & 1 & 1 & 1 & 0 & 0 & 0 & 0\tabularnewline
\hline 
\multirow{4}{*}{Comments} & Attended & 3 & 5 & 0 & 1 & 2 & 3 & 2 & 1\tabularnewline
 & Didn't attend & 0 & 3 & 3 & 1 & 2 & 1 & 3 & 2\tabularnewline
 & Unresponsive & 0 & 2 & 0 & 0 & 0 & 3 & 0 & 0\tabularnewline
 & Dropped out & 0 & 1 & 0 & 1 & 0 & 2 & 0 & 0\tabularnewline
\hline 
\multirow{2}{*}{Conference} & Attended & 3 & 8 & 3 & 2 & 4 & 4 & 4 & 3\tabularnewline
 & Dropped out & 0 & 0 & 0 & 0 & 0 & 0 & 1 & 0\tabularnewline
\hline 
\end{tabular}\end{footnotesize}
\par\end{center}

\dataavailability{Data supporting the findings of this study that
is not already contained within the article or derived from listed
public domain resources are available on request from the corresponding
author. This data is not publicly available due to ethical restrictions
based on the nature of this work.}

\authorcontribution{MOA conceived the programme and its evaluation,
performed the analysis, and wrote the paper.}

\competinginterests{The author declares that they have no conflict
of interest.}
\begin{acknowledgements}
We thank Dominic Galliano, Olivia Keenan, Charlotte Thorley, and Jennifer
DeWitt for helpful discussions. MOA holds a UKRI (STFC / EPSRC) Stephen
Hawking Fellowship EP/T01735X/1 and received funding from the Ogden
Trust. This programme has been supported by a QMUL Centre for Public
Engagement Large Award, and STFC Public Engagement Small Award ST/N005457/1.
\end{acknowledgements}
\DeclareRobustCommand{\disambiguate}[3]{#1}\bibliographystyle{copernicus}
\bibliography{prise2020}

\end{document}